\documentclass{aa} 
\usepackage{txfonts}
\usepackage{longtable}  
\usepackage{rotating}
\usepackage{natbib}
\usepackage{graphicx}
\usepackage{graphics}
\usepackage{psfrag}
\usepackage{amssymb}
\bibpunct{(}{)}{;}{a}{}{,}
\def\Teff{$T_{\mathrm{eff}}$}
\def\logg{\ensuremath{\log g}}
\def\vmic{$\upsilon_{\mathrm{mic}}$}

\def\vsini{\ensuremath{{\upsilon}\sin i}}
\def\kms{$\mathrm{km\,s}^{-1}$}

\def\loggf{log$gf$}
\def\vr{${\upsilon}_{\mathrm{r}}$}

\def\nlte{non--LTE}
\def\llm{{\sc LLmodels}}

\begin{document} 
\title{Abundance analysis of seven $\delta$~Scuti Stars} 
\subtitle{} 
\author{L. Fossati\inst{1}       \and 
        K. Kolenberg\inst{1,2}	 \and
	P. Reegen\inst{1}	 \and
	W. Weiss\inst{1}} 
\offprints{L.~Fossati} 
\institute{
	Institut fur Astronomie, Universit\"{a}t Wien, 
	T\"{u}rkenschanzstra{\ss}e 17, 1180 Wien, Austria.\\
	\email{fossati@astro.univie.ac.at; reegen@atro.univie.ac.at; weiss@astro.univie.ac.at} 
	\and 
	Instituut voor Sterrenkunde, Celestijnenlaan 200D, 3001 Leuven,
	Belgium. 
	\email{kolenberg@astro.univie.ac.at}
} 
\date{} 
\abstract 
{The current knowledge of the abundance pattern in $\delta$~Scuti stars is
based on the analysis of just a few field stars.}
{We aim to determine the general chemical properties of the atmospheres of
$\delta$~Scuti stars based on a statistically relevant sample of
stars and will investigate whether the abundance pattern is close to 
solar, an assumption generally made for pulsation models.} 
{We have analysed high--resolution, high signal--to--noise ratio spectra of 
seven field $\delta$~Scuti stars. We derived the fundamental parameters and the
photospheric abundances and compared them to a similar sample of cluster
$\delta$~Scuti stars.} 
{With the use of a $t$--test we demonstrated that there is no difference 
between the two samples, which allows us to merge them, resulting in a sample 
of fifteen $\delta$~Scuti stars. We did not find any substantial difference 
between the abundance pattern of our sample of $\delta$~Scuti stars and a 
sample of normal early A-- and late F--type stars. One field star 
in our sample, HD~124953, is most likely a pulsating Am star.} 
{} 
\maketitle
\titlerunning{Properties of Delta Scuti Stars}
\authorrunning{L.~Fossati et al.}
\section{Introduction}\label{introduction}
It is widely accepted that chemical peculiarity and pulsation are 
almost mutually exclusive. The diffusion hypothesis \citep{vauclair} explains
in details why this occurs. 
$\delta$~Scuti pulsation is driven by the $\kappa$ mechanism in the 
\ion{He}{ii} ionisation zone. For stars in which helium is gravitationally 
settled due to diffusion the driving mechanism is less efficient and 
pulsation is inhibited. Nevertheless, there are a few chemically peculiar Am 
and Ap stars that do pulsate. It is thought that in these stars, the pulsation 
is sufficiently laminar not to disrupt the anomalous distribution of 
elements. 

Since the \ion{He}{ii} ionisation zone drives the pulsation it is expected 
that the photospheric abundances of the individual elements have a negligible 
influence on the pulsation properties.  
In other words, the specific 
``skin diseases" \citep{breger2006} of some $\delta$~Scuti stars may not be 
relevant for modelling their pulsations. 
However, it is only through detailed analyses that we can quantify the 
element abundances (diagnose the skin disease, to stay with the analogon).

Throughout the literature, only a few detailed abundance analyses of 
$\delta$~Scuti stars can be found 
\citep[e.g.][]{mittermayer2003,zima2007,bruntt}. 
In this paper, we present one of the few detailed element abundance analyses 
available for $\delta$~Scuti stars and the first of a statistically relevant
sample of stars. We carried out spectroscopic abundance analyses for seven 
field $\delta$~Scuti stars. We want to answer the question whether the 
abundance pattern of field $\delta$~Scuti stars is substantially different from 
those of cluster $\delta$~Scuti stars, such as the stars in the Praesepe 
cluster analysed by \citet{fossati2007} and \citet{fossati2008}, and how the 
abundance pattern of $\delta$~Scuti stars compares with those of normal F-- and 
A--type stars. 

The current analysis is part of a large project carried out by the authors 
and several other collaborators that aims at the analysis of chemical 
\citep[see e.g.][]{fossati2008} and magnetic 
\citep[see ][]{stefano2006,john2006,john2007} properties of early--type stars 
in open clusters. 

\citet{breger2000} pointed out that ``about 50\% of all main sequence stars
inside the instability strip are $\delta$~Scuti pulsators" and that probably 
the other 50\% ``do pulsate with amplitudes below the present level of 
detection". Within this context our analysis turns out to be also an abundance 
analysis of non--chemically peculiar (CP: Fm, Am, $\lambda$~Boo 
and Ap stars) early F-- and late A--type field stars. 

The observed stars, adopted instrumentation and the abundance analysis procedure
are described in Sect.~\ref{observations}. In Sect.~\ref{results} we present our
results including a description of the pulsation characteristics for each star
of our sample. Discussion and conclusion are given in Sect.~\ref{discussion} and
Sect.~\ref{conclusions}, respectively.
\section{Observations, data reduction and abundance analysis}\label{observations}
We observed seven $\delta$~Scuti stars, present in the catalogue of
\citet{rodriguez} with the SOPHIE spectrograph at the Observatoire de 
Haute--Provence (OHP) from 2007 March 10--12. 
SOPHIE is a cross--dispersed \'{e}chelle spectrograph mounted on the 1.93--m 
telescope at the OHP. The spectrograph is fed from the Cassegrain focus 
through pairs of optical fibers, one of which is used for starlight and the 
other can be used for either the wavelength calibration lamp or the sky 
background, but can also be masked. The spectra cover the wavelength range 
3872--6943~\AA\, and the instrument allows either mid--resolution mode 
(R$\sim$40000) or high--resolution mode (R$\sim$75000). All the stars of our 
sample were observed in high resolution.
The spectra were automatically reduced using a pipeline adapted from the HARPS 
software designed by Geneva Observatory (see the SOPHIE web page
\footnote{\tt www.obs-hp.fr/www/guide/sophie/sophie-eng.html}).

The sample of stars observed and analysed in this paper are listed in
Table~\ref{tabella radec}. Three stars are F--type, while the other four are
A--type stars.
\begin{table*}[ht]
\caption[ ]{Basic data for the observations and the pulsation properties of the program stars.}
\label{tabella radec}
\begin{center}
\begin{tabular}{rrrrrrrrl}
\hline
\hline
HD & HJD & M$_{\it{v}}$ & Spectral Type & SNR & Exp. Time   & log $P$ & $Q$   & Remarks \\
   &	 & [mag]	&		&     &[s]	    &	      & [d]   &         \\
\hline
127929 & 2454170.650 & 6.27 & F0III	& 238 & 550	    & $-$1.06   & 0.018           \\
138918 & 2454170.693 & 3.80 & F0IV	& 286 & 550	    & $-$0.81   & 0.032 & binary  \\
143466 & 2454171.703 & 4.97 & F0IV	& 465 & 550	    & $-$1.12   & 0.029           \\
124675 & 2454170.598 & 4.74 & A8IV	& 390 & 600	    & $-$1.19   & 0.015 & binary  \\
124953 & 2454170.614 & 5.97 & A8III	& 257 & 550	    &	        &       & binary? \\
125161 & 2454170.626 & 4.75 & A9V	& 295 & 550	    & $-$1.58   & 0.013           \\
127762 & 2454170.686 & 3.00 & A7III	& 334 & 550	    & $-$1.14   & 0.017 & binary  \\
\hline
\end{tabular}
\end{center}
\smallskip 

The SNR is calculated at $\sim$5500~\AA\ in a bin of 0.5 \AA. The exposure time is in 
seconds. The HJD indicates the Heliocentric Julian Date at the end of the 
exposure. The analysed spectrum of HD~127929 is the sum of two spectra with the
same exposure time. Periods and $Q$ values are taken from \citet{rodriguez} and 
\citet{lopez}.
\end{table*}

Almost all the stars in the sample have a high projected rotational 
velocity (\vsini\, $\geq$ 50 \kms). For these objects the continuum 
normalisation is crucial and it was performed without the use of any automatic
continuum fitting process. First we divided the spectra in portions of about
700~\AA\, each. Subsequently, we performed the normalisation by fitting a spline to
carefully selected continuum windows. These were identified by comparison with a
synthetic spectrum of the same approximate \Teff, \logg\, and \vsini. It was 
not possible to determine a correct continuum level at wavelengths shorter 
than the H$\gamma$ line (4340~\AA) since there are not enough continuum 
windows in the spectra because of the crowding of spectral lines in this region.

Model atmospheres were calculated with \llm, an LTE code which uses direct
sampling of the line opacities \citep{denis2004} and allows to compute models 
with an individualised abundance pattern. Spectral line data were extracted 
from the VALD database \citep{vald1,vald2,vald3}.

\begin{figure}[ht]
\begin{center}
\resizebox{\hsize}{!}{\includegraphics{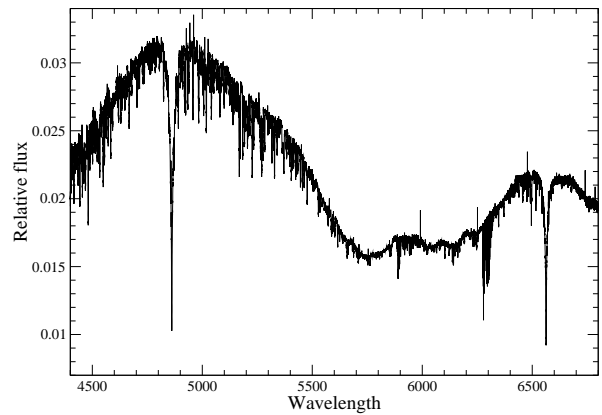}}
\caption{The spectrum of HD~124953 from the SOPHIE reduction pipeline.}
\label{sophie}
\end{center}
\end{figure}

\begin{figure}[ht]
\begin{center}
\resizebox{\hsize}{!}{\includegraphics{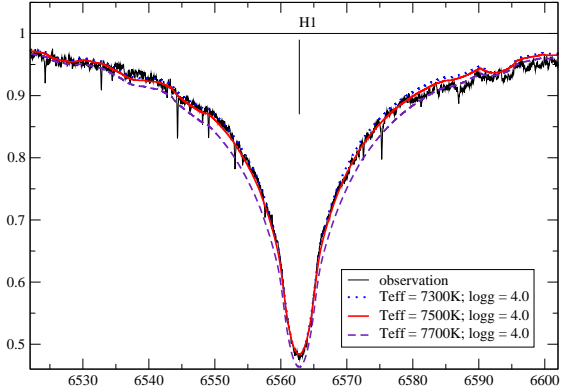}}
\caption{H$\alpha$ line fitting for HD~124675 (\vsini\, = 121 \kms) to 
estimate the effective temperature. The three plotted synthetic spectra 
correspond to \Teff\, of 7300~K (dotted line), 7500~K (full line) and 
7700~K (dashed line).}
\label{hline}
\end{center}
\end{figure}

The fundamental parameters were derived spectroscopically using as starting
models those from Str\"omgren and Geneva
photometry, adopting calibrations from \citet{napiwotzki} and
\citet{geneva_cal}. Effective temperature and \logg\, were obtained from 
\ion{Fe}{i} and \ion{Fe}{ii} lines and from fitting the H$\alpha$ line 
wings. For most of the \'echelle spectrographs it is very difficult 
to use the Balmer lines to constrain the fundamental parameters, since the
orders are not large enough to completely include the wings.
With the SOPHIE spectrograph, however, it is possible due to the reduction 
pipeline which rescales and merges the orders before the normalisation. In 
this way, a single spectrum is obtained, (shown in Fig.~\ref{sophie}) that 
allows the normalisation of the Balmer lines. We adopted H$\alpha$ 
for the determination of \Teff\, since it has several continuum points 
close to the wings. An example of the fit to an H$\alpha$ line is given in
Fig.~\ref{hline}. The plot demonstrates that we can determine \Teff\, with an
uncertainty of about 200~K. This method is only weakly dependent on the
projected rotational velocity (\vsini). In most cases the photometric and 
spectroscopic temperatures were consistent. \citet{bruntt} showed that the 
uncertainty on the spectroscopically derived \logg\, value is dependent on 
\vsini. Their Table~3 shows the increment of the uncertainty on \logg\, with 
increasing \vsini\, from 10~\kms\, to 60~\kms. A linear extrapolation of this 
uncertainty at \vsini\, = 110~\kms\, (about the mean \vsini\, in our sample) 
leads to an error bar on \logg\, of about 0.22~dex. We derived \vmic\, as 
in \citet{fossati2008} and assumed the same uncertainty. Therefore the 
uncertainties on \Teff, \logg\, and \vmic\, are estimated to be 200~K, 
0.22~dex and 0.7~\kms, respectively.

\begin{figure}[ht]
\begin{center}
\resizebox{\hsize}{!}{\includegraphics{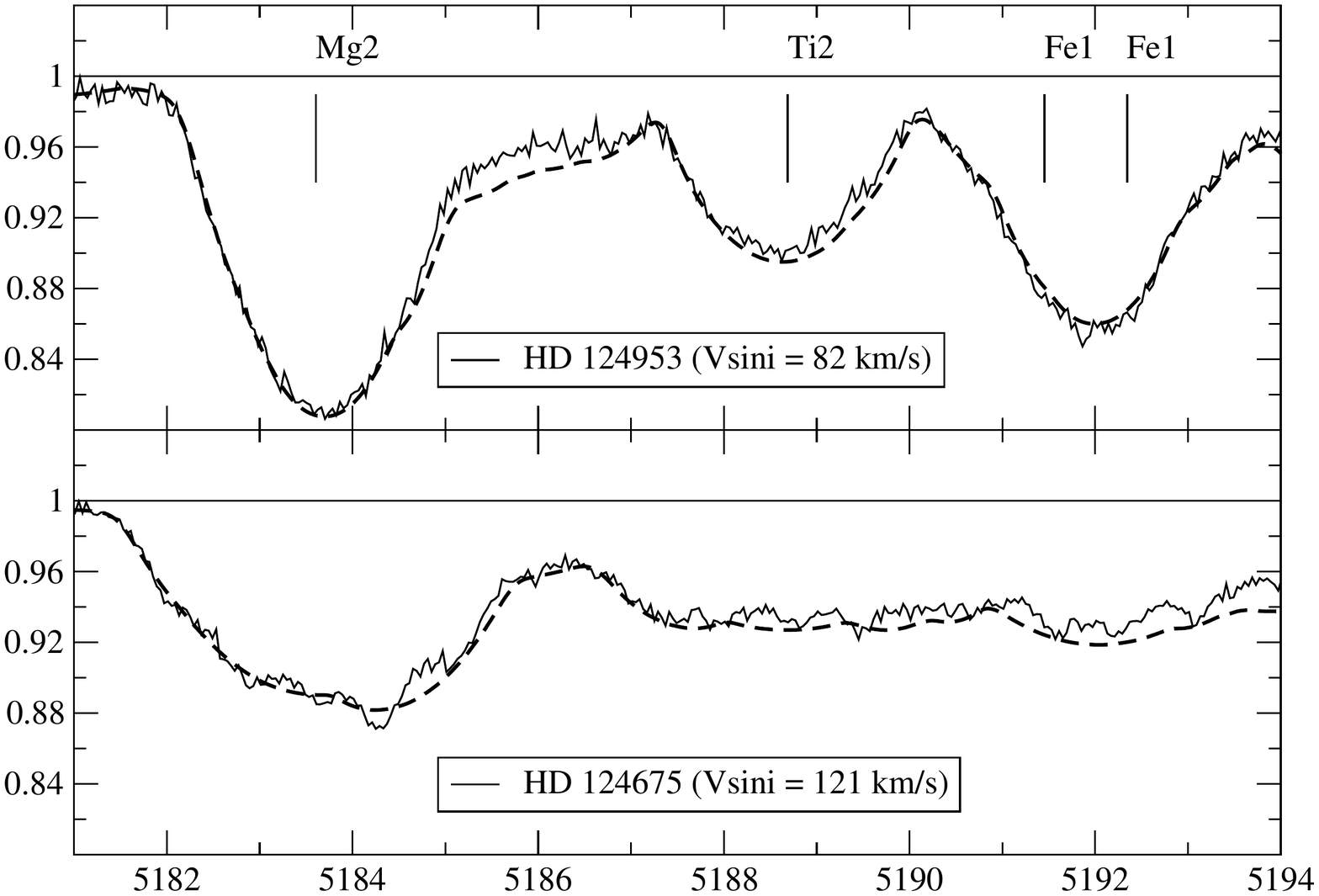}}
\resizebox{\hsize}{!}{\includegraphics{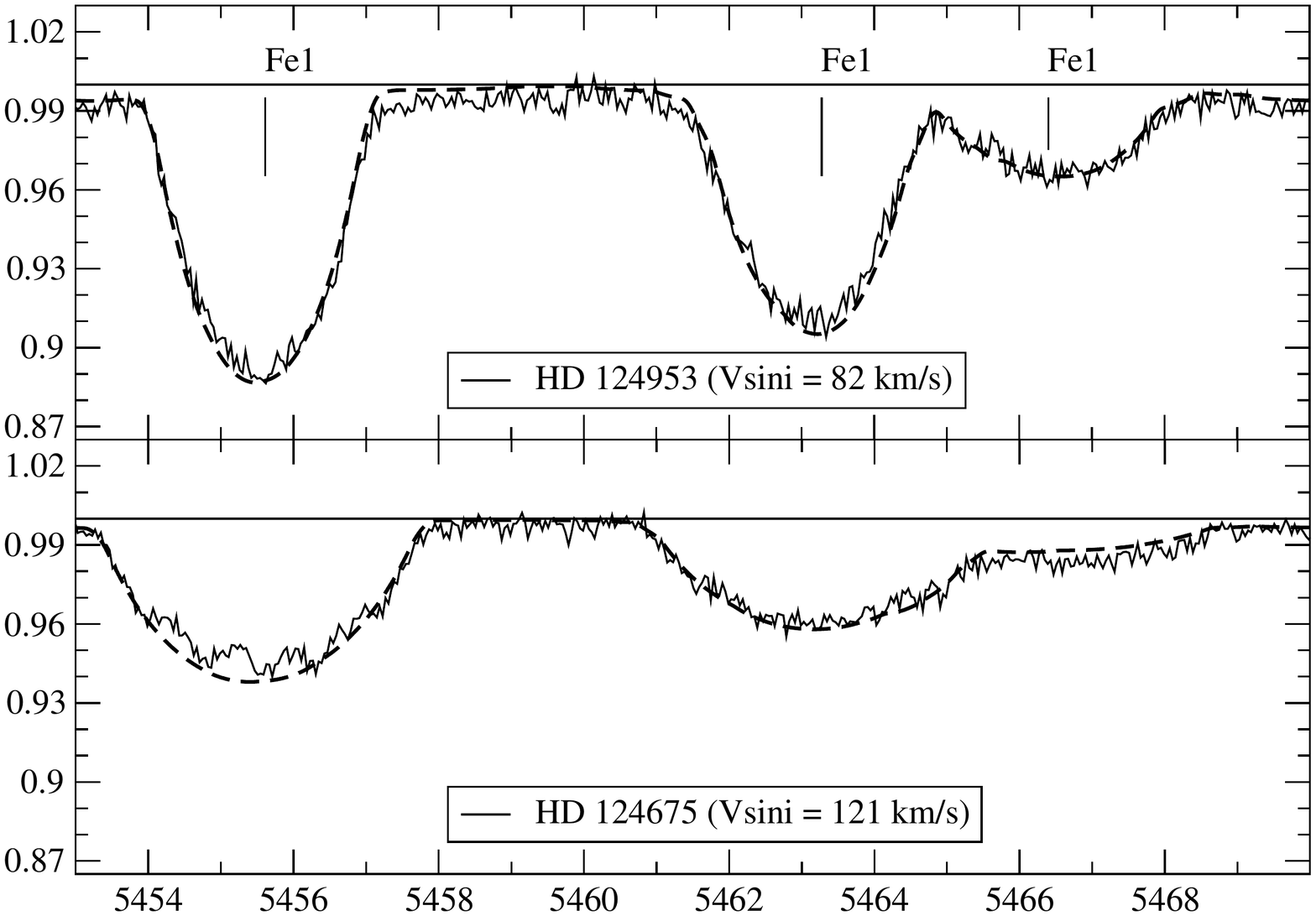}}
\caption{Portions of the observed spectra of HD~124953 (\vsini\, = 82 \kms, 
upper panels) and of HD~124675 (\vsini\, = 121 \kms, lower panels) and the 
synthetic spectra are shown with full and dashed lines, respectively.}
\label{spec_short}
\end{center}
\end{figure}

An example of the observed spectra of the moderate rotator HD~124953 
(\vsini\, = 82 \kms) and the fast rotator HD~124675 (\vsini\, = 121 \kms) 
and the synthetic spectra obtained from the abundance analysis is 
shown in Fig.~\ref{spec_short}.

Details on the parameter determination and on the abundance analysis method
including a thorough discussion of the abundance uncertainties for fast rotating 
stars are presented in \citet{fossati2007} and \citet{fossati2008}.

We adopted the results by \citet{fossati2008} for the estimation of the 
abundance uncertainties, which are affected by fast rotation of the stars and 
the uncertainties on \Teff\, and \vmic. For the stars in our sample the 
uncertainty is about 0.25~dex.

Another source of uncertainty, which is dependent on \vsini\, is the 
continuum normalisation. To quantify this uncertainty we performed the 
following test. We derived the abundance of the \ion{Fe}{ii} line at 
5325.6~\AA\, for HD~124953 (\vsini\, = 82 \kms) with the adopted normalised 
observed spectrum and with the spectrum multiplied/divided by 0.99. In this 
way we increased/decreased the continuum level by 1\%, which we estimate 
to be a reasonable uncertainty. The difference between the abundances obtained 
in this way is about 0.1~dex, consistent with what was obtained with the same
experiment by \citet{fossati2008}. Their conclusion is that the abundance error
bar due to the uncertainty on the continuum normalisation is increasing with
\vsini\, up to 0.2~dex for stars with \vsini\, $\sim$ 200~\kms. We believe 
that in most cases this error bar is an upper limit since the line--by--line 
abundance analysis method, used in this work, allows a very careful line 
selection that rejects all the lines for which the continuum level looks 
uncertain. Under the assumption that no systematic errors are present 
the uncertainty will decrease when several lines are used for an element.
For a more thorough evaluation see \citet{erspamer} who showed how this source 
of error depends on the element. 

The photometric and the adopted spectroscopic fundamental parameters are shown
in Table~\ref{parameters}.
\begin{table}[ht]
\caption[ ]{Atmospheric fundamental parameters obtained from Str\"{o}mgren and Geneva photometry 
for the sample stars and their derived radial velocity.}
\label{parameters}
\begin{center}
\begin{tabular}{rrrrrr}
\hline
\hline
 & \multicolumn{2}{c}{Str\"{o}mgren} & \multicolumn{2}{c}{Geneva} & \\
\hline
HD & \Teff & \logg  & \Teff & \logg & \vr    \\
   &   [K] & [cgs]  &  [K]  & [cgs] & [\kms] \\
\hline
127929 & 7552 & 3.67 & 7361 & 3.94 & $-$18.9 \\
138918 &      &      & 7345 & 4.02 & $-$41.6 \\
143466 & 7380 & 3.96 &      &      & $-$13.4 \\
124675 &      &      & 7470 & 4.08 & $-$17.1 \\
124953 &      &      & 7388 & 4.38 & $+$5.3  \\
125161 & 7832 & 4.14 & 7755 & 4.46 & $-$16.9 \\
127762 & 7797 & 3.67 & 7581 & 3.97 & $-$31.6 \\
\hline
\end{tabular}
\end{center}
\smallskip 

 The \vr, in \kms, has an error bar of about 4~\kms.
\end{table}


We determined the abundance by fitting the core of the selected 
lines. The spectra were synthesised with Synth3 \citep{synth3}. The fitting 
procedure and the determination of the abundances was done in an iterative 
process in order to get better values for the abundances of the more blended 
lines. The error bar associated with each 
element is the standard deviation from the mean abundance of the selected 
lines of that element. For all the elements that were not analysed we 
adopted the solar abundance from \citet{asplundetal2005}. 

HD~138918, HD~124675, HD~127762 and possibly HD~124953 are binary stars 
\citep{hoffleit1964}. No direct indication of the spectrum of 
the secondary star was found in our data, since its flux is probably below 
5\% of the primary's flux. 

We measured the radial velocity (\vr) and \vsini\, from fitting the observed 
data with synthetic spectra. This analysis was based on about eighty lines and the
typical uncertainty on \vr\, and \vsini\, is of the order of 4 \kms\, and 5\%, 
respectively.
\section{Results}\label{results}
In this section we present the results of the abundance analysis obtained for
each star in our sample. The pulsation characteristics of the individual 
targets are also briefly mentioned. Due to the brightness of most of 
the targets their photometric variation has not always been well studied, due
to a lack of good comparison stars in the vicinity. Hence the information 
on the pulsation may not be fully reliable. 

\begin{figure*}[ht]
\begin{center}
\rotatebox{270}{
\includegraphics[width=230mm]{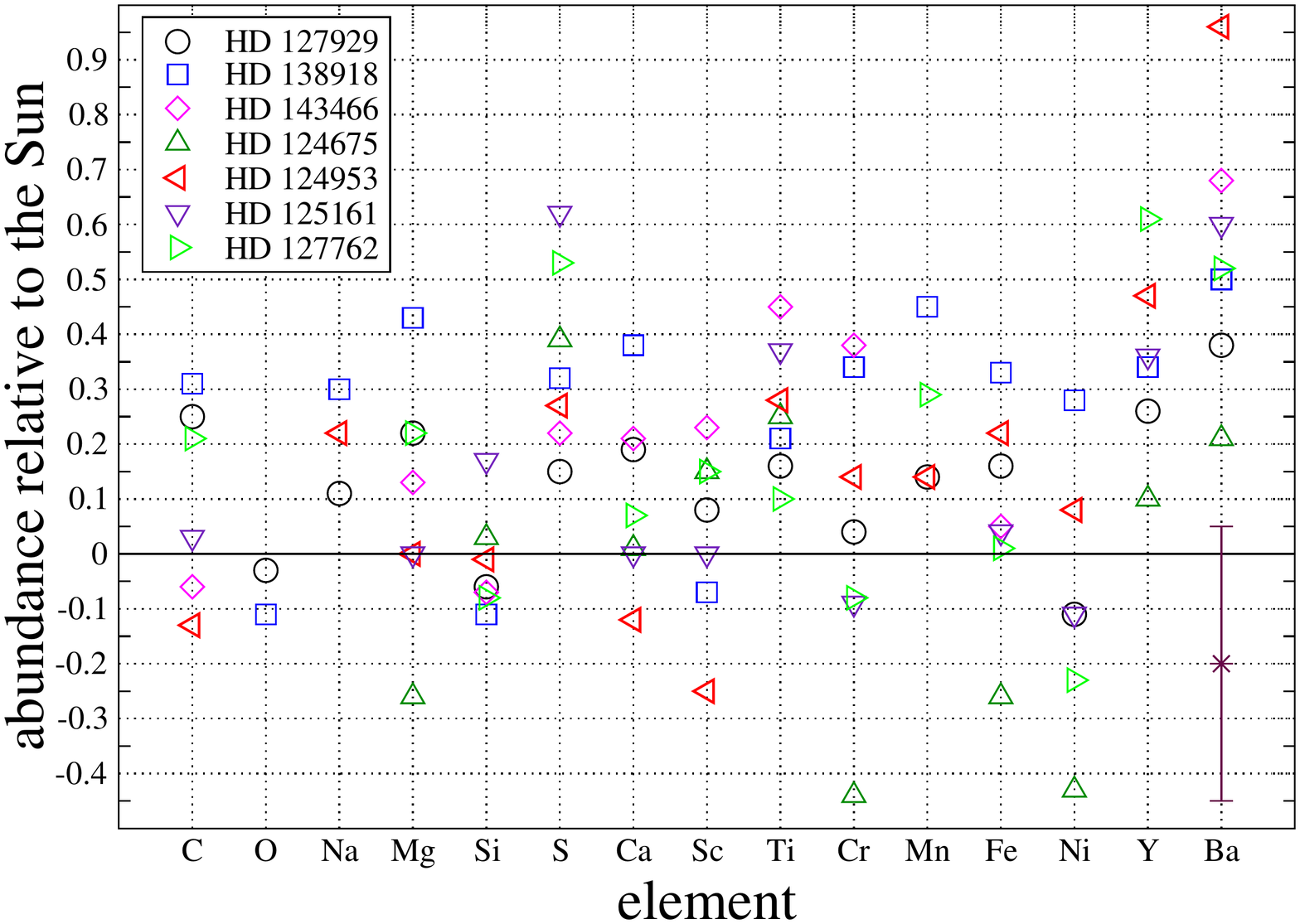}
}
\caption{Elemental abundances relative to the Sun for the program stars. 
The solar abundances are taken from \citet{asplundetal2005}. In order to 
show a more readable Figure, the error bars (Table~\ref{abundance_table}) 
are omitted. The abundances of the Am--$\delta$~Scuti star are highlighted
in red. The error bar associated to the star symbol (bottom right of the
Figure) shows the typical uncertainty associated with the plotted abundances.}
\label{abundance_figure}
\end{center}
\end{figure*}

The derived abundances and the estimated uncertainties are given in 
Table~\ref{abundance_table} and shown online in Fig.~\ref{abundance_figure}.
\begin{table*}[ht]
\caption[ ]{Abundances ($\log(N_{X}/N_{\rm tot})$) of the program stars. }
\label{abundance_table}
\begin{center}
\scriptsize
\begin{tabular}{rrrrrrrrrr}
\hline
\hline
\multicolumn{2}{c}{ } & \multicolumn{7}{c}{$\delta$~Scuti stars} & Solar \\
At.N.&Element   & HD~127929    & HD~138918    & HD~143466    & HD~124675    & HD~124953    & HD~125161    & HD~127762  & Abundances \\
\hline
6 & C	        &$-3.40(-;1)  $&$-3.34(09;2) $&$-3.71(-;1)  $&	            &$-3.78(06;2) $&$-3.62(09;2) $&$-3.44(-;1)  $&$-3.65 $\\  
8 & O	        &$-3.41(-;1)  $&$-3.49(-;1)  $&	             &	            &	           & 	          &	         &$-3.38 $\\  
11& Na          &$-5.76(06;2) $&$-5.57(10;2) $&	             &	            &$-5.65(01;2) $& 	          &	         &$-5.87 $\\  
12& Mg          &$-4.29(04;4) $&$-4.08(06;4) $&$-4.38(-;1)  $&$-4.77(11;3) $&$-4.51(10;5) $&$-4.39(09;2) $&$-4.29(01;2) $&$-4.51 $\\  
14& Si          &$-4.59(-;1)  $&$-4.64(-;1)  $&$-4.60(-;1)  $&$-4.50(-;1)  $&$-4.54(-;1)  $&$-4.36(-;1)  $&$-4.61(-;1)  $&$-4.53 $\\  
16& S	        &$-4.75(03;2) $&$-4.58(06;2) $&$-4.68(08;2) $&$-4.51(-;1)  $&$-4.63(07;5) $&$-4.28(05;2) $&$-4.37(04;2) $&$-4.90 $\\  
20& Ca          &$-5.54(11;7) $&$-5.35(14;9) $&$-5.52(01;2) $&$-5.72(-;1)  $&$-5.85(09;5) $&$-5.73(07;6) $&$-5.66(10;5) $&$-5.73 $\\  
21& Sc          &$-8.91(11;3) $&$-9.09(04;2) $&$-8.76(04;2) $&$-8.84(15;3) $&$-9.24(13;4) $&$-8.99(01;2) $&$-8.84(07;3) $&$-8.99 $\\  
22& Ti          &$-6.98(11;4) $&$-6.93(13;7) $&$-6.69(13;2) $&$-6.89(01;2) $&$-6.86(13;7) $&$-6.77(11;7) $&$-7.04(09;4) $&$-7.14 $\\  
24& Cr          &$-6.36(07;3) $&$-6.06(12;3) $&$-6.02(-;1)  $&$-6.84(04;3) $&$-6.26(23;4) $&$-6.49(19;5) $&$-6.48(09;3) $&$-6.40 $\\  
25& Mn          &$-6.51(-;1)  $&$-6.20(09;2) $&	             &	            &$-6.51(05;6) $& 	          &$-6.36(-;1)  $&$-6.65 $\\  
26& \ion{Fe}{i} &$-4.44(05;26)$&$-4.26(10;40)$&$-4.54(06;16)$&$-4.85(11;27)$&$-4.38(08;65)$&$-4.56(08;31)$&$-4.58(11;29)$&        \\
26& \ion{Fe}{ii}&$-4.39(05;5) $&$-4.24(14;3) $&$-4.56(-;1)  $&$-4.71(-;1)  $&$-4.35(15;3) $&$-4.55(07;5) $&$-4.55(12;5) $&        \\
26& Fe          &$-4.43(06;31)$&$-4.26(10;43)$&$-4.54(06;17)$&$-4.85(12;28)$&$-4.37(08;68)$&$-4.55(08;36)$&$-4.58(11;34)$&$-4.59 $\\  
28& Ni          &$-5.92(11;2) $&$-5.53(11;4) $&	             &$-6.24(-;1)  $&$-5.73(09;5) $&$-5.92(07;3) $&$-6.04(08;2) $&$-5.81 $\\  
39& Y	        &$-9.57(-;1)  $&$-9.49(11;2) $&	             &$-9.73(-;1)  $&$-9.36(-;1)  $&$-9.47(-;1)  $&$-9.22(-;1)  $&$-9.83 $\\  
56& Ba          &$-9.49(-;1)  $&$-9.37(-;1)  $&$-9.19(-;1)  $&$-9.66(-;1)  $&$-8.91(-;1)  $&$-9.27(-;1)  $&$-9.35(-;1)  $&$-9.87 $\\  
\hline
&   \Teff       & 7600	       & 7800         & 7380	     & 7500	    & 7600         & 7700	  & 7800	 & \\
&   \logg       & 3.70	       & 3.50         & 3.90	     & 4.00	    & 4.10         & 4.45	  & 3.75	 & \\
&   \vmic       & 2.7	       & 2.6          & 2.7	     & 2.7	    & 2.6          & 2.6	  & 2.6 	 & \\
&   \vsini      & 73	       & 88           & 136	     & 121	    & 82	   & 131	  & 121 	 & \\
&$\sigma_{\rm abn}$& 0.23      & 0.23         & 0.26	     & 0.25         & 0.23	   & 0.26	  & 0.25	 & \\
\hline
\end{tabular}
\end{center}
\smallskip 

The estimated internal errors in units of 0.01 dex and the number of selected 
lines are given in parentheses. Abundances obtained from one line have 
no error ($-$). The uncertainties on the fundamental parameters, \Teff, \logg, 
and \vmic\, are 200~K, 0.22~dex and 0.7~\kms. The uncertainty on \vsini\, is 5\%.
The estimated uncertainty on the abundances ($\sigma_{\rm abn}$), including 
contributions from the high \vsini\, and uncertainties on \Teff\, and \vmic, 
are given in the last row of the Table.
\end{table*}

The abundances of the whole sample of field $\delta$~Scuti stars appear to be 
quite dispersed, except for e.g. Si, Ti and Mn. The 
abundances of most of the elements (C, O, Mg, Si, Ca, Sc, Cr, Fe and Ni) 
appear to be close to solar, while Na, S, Ti, Mn, Y and Ba show a slight (but 
clear) overabundance.
\subsection{HD~127929}\label{hd127929}
The $\delta$~Scuti nature of HD~127929 (HR~5437, ER Dra) was discovered by 
\citet{jiang1988}. The star has a pulsation amplitude of about 0.02~mag 
in $V$. \citet{li1992} found the main pulsational frequency to be 
11.38993 c/d. This frequency, combined with another frequency at 8.5367 c/d 
found by \citet{paparo1990} suggests that HD~127929 is a $\delta$~Scuti star 
pulsating in the radial fundamental and radial first overtone modes. 

The analysed spectrum of this star is the sum of two consecutive spectra with 
equal exposure time. This was done to increase the signal--to--noise ratio 
and to avoid saturating the CCD. The abundances determined for this star 
seem to be almost solar for all elements, except for a slight Ba 
overabundance.
\subsection{HD~138918}\label{hd138918}
HD~138918 (HR~5789, $\delta$~Ser) is part of a multiple system consisting of a 
close binary \citep{muller1950} with 
a separation angle of 4.000$\pm$0.001 arcsec \citep{alzner1998,prieur2002}, 
and two other stars of magnitude 13 and 14 in $V$. The primary is an A9V star with 
4.25 mag in $V$ ($\delta$~Ser) and its companion an A7V star with 5.2 mag in $V$ 
\citep{baizepetit89}.

\citet{lopez1987} found two main frequencies at 6.4227 c/d and
7.8869 c/d, which were identified as the first and second overtone, with a
ratio of 0.814.
The star is one of the brightest $\delta$~Scuti stars in the sky
\citep{breger2008}, which will be suitable for photometric monitoring with 
the BRITE microsatellite \citep{kaiser2008}. It has a pulsation amplitude of 
about 0.04 mag in $V$. 

\begin{figure}[ht]
\begin{center}
\resizebox{\hsize}{!}{\includegraphics{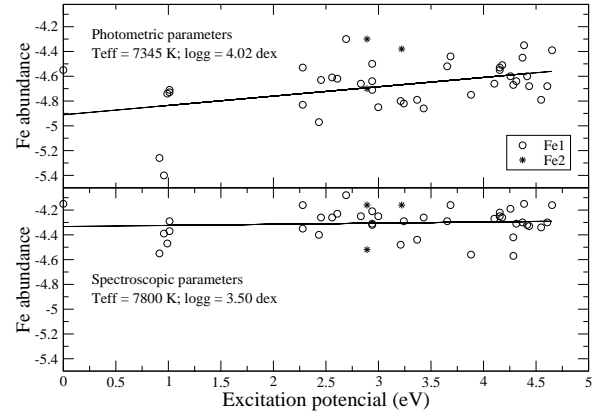}}
\caption{Correlation between \ion{Fe}{i} (circles) and \ion{Fe}{ii} 
(stars) abundances and excitation potencial adopting photometric (upper panel) and
spectroscopic (lower panel) parameters for HD~138918.}
\label{temperature}
\end{center}
\end{figure}

There is a large difference between the parameters
obtained from photometry (\Teff\, = 7345 K; \logg\, = 4.02) and the adopted
spectroscopic parameters (\Teff\, = 7800 K; \logg\, = 3.50). 
Figure~\ref{temperature} shows the correlation with excitation potencial of the
abundances of \ion{Fe}{i} (circles) and \ion{Fe}{ii} (stars) lines adopting 
the fundamental parameters obtained from photometry (upper panel) and from 
spectroscopy (lower panel). The positive trend in the upper panel of 
Fig.~\ref{temperature} is mainly due to the presence of two lines with low
excitation potencial, which are very sensitive to changes in
temperature\footnote{Note that the scatter of the points is significantly 
lower for the higher temperature as determined from spectroscopy.}. 

We are not able to explain the reason for the discrepancy between the parameters
derived photometrically and spectroscopically. It could be that this 
difference is due to the effects of the secondary, of which we do not see 
any lines, but this is just a speculation. 

The analysis of this star yields a general overabundance of about 0.3 dex 
for all the elements, while O, Si and Sc are consistent with the solar
abundance.
\subsection{HD~143466}\label{hd143466}
The variability of HD~143466 (HR~5960, CL~Dra) was discovered by \citet{breger1969}, who 
classified the star as a $\delta$~Scuti variable. \citet{dupuy1983} 
found a main frequency $f_1$ at 14.7929 c/d (which could also be 14.5985 c/d) 
and a secondary frequency $f_2$ at 20.2840 c/d. The star has a pulsation 
amplitude of about 0.01 mag in $V$. 

Due to the rather high rotational velocity of the star (\vsini\, = 136 \kms) it
was only possible to derive abundances of the main elements.
Except for an overabundance of Ti, Cr and Ba, the other elements show
almost solar abundances.
\subsection{HD~124675}\label{hd124675}
HD~124675 (HR~5329, $\kappa^2$~Boo) is a binary star and one component 
is known to be a $\delta$~Scuti pulsator of spectral type A8IV. The other 
component ($\kappa^1$~Boo) is an F1V star. 
\citet{bakos1986} determined the period of the system to be 1790 days, and the 
semiamplitude of the velocity curve is 7.2~\kms. 
\citet{frandsen1995} 
found that the star shows $\delta$~Scuti--type pulsation with a few closely
spaced modes. 
They found frequencies at $f_1 = 15.43$ c/d, 
$f_2 = 15.58$ c/d,
$f_3 = 14.52$ c/d, 
$f_4 = 15.81$ c/d. 
The star has a pulsation amplitude (sum of all modes) of about 
0.05 mag in $V$. Their preliminary mode identification points towards the 
second
overtone for the $f_1$ mode ($Q \simeq 0.016$).
The other modes may perhaps have $\ell$ values of 2 and 1, but more 
observations and detailed modelling including the effects of rotation are 
needed to obtain a robust mode identification. 
\citet{walker1987} and \citet{kennelly1991} showed from line profile analysis 
that the star has non--radial pulsation modes. Thanks to the fast rotation of 
the star it is possible to detect modes with even high degrees. 
\citet{kennelly1991} found that a mode with high $\ell$ ($\simeq 12$) in 
combination with a low--degree non--radial mode can explain the observed line 
profile variations. From high--quality data more non--radial pulsation modes 
could likely be found.

The abundances determined for HD~124675 are almost solar for all the elements, 
except of S (overabundant) and Cr and Ni (underabundant). Among the stars of 
our sample, HD~124675 shows the lowest average abundance.
\subsection{HD~124953}\label{hd124953}
HD~124953 (HR~5343, CN Boo) was discovered to show $\delta$~Scuti pulsation 
by \citet{costa1979}. They found a frequency of about 25 c/d which 
they identified through the $Q$ value of 0.025 as the first overtone radial 
mode. Given the pulsation, they concluded that the classification 
of HD~124953 as an Am star \citep{hoffleit1964} must be wrong, since Am stars 
are not supposed to pulsate (see Sect.~\ref{disc-hd124953}). The star has a 
pulsation amplitude of about 0.03 mag in $V$. In the Bright Star Catalogue 
\citep{hoffleit1964} the star is marked as a suspected spectroscopic binary.
Our spectrum does not show any feature that indicate this.

The error bar associated with the Cr abundance is quite large relative to the 
other elements. This is due to the difference in abundance found for 
\ion{Cr}{i} and \ion{Cr}{ii} (we used two spectral lines for each ionisation 
stage). In particular one \ion{Cr}{ii} line ($\sim$5280~\AA) deviates from the 
mean abundance. We checked the \loggf\, value of this line by comparing 
with a spectrum of the Sun, and it seems to be correct. An improper continuum 
normalisation could explain the discrepancy for this line.

HD~124953 shows a slight underabundance of C and Ca, an underabundance of Sc, a
slight overabundance of the Fe--peak elements and a clear overabundance of Y 
and Ba. The other elements show almost solar abundances. This abundance 
pattern, typical of Am stars, will be discussed in detail in 
Sect.~\ref{disc-hd124953}.
\subsection{HD~125161}\label{hd125161}
HD~125161 (HR~5350, $\iota$ Boo) was discovered to be a $\delta$~Scuti star by 
\citet{albert1980}. It was discussed by several authors 
\citep{kiss1999,zhou1999} showing pulsation in photometry, which can be 
fitted with one single period. \citet{zhou1999} found the main frequency to be 
$f_1$=37.6804 c/d and also found a low--amplitude secondary peak at 
$f_2$=36.4111 c/d. The star has a pulsation amplitude of about 0.01 mag in $V$. 

Only S, Ti, Y and Ba appear to be overabundant, while all the other
elements show solar abundances. As for HD~124953 (see
Sect.~\ref{hd124953}), the error bar associated with the Cr abundance is 
large and for the same reason.
\subsection{HD~127762}\label{hd127762}
HD~127762 (HR~5435, $\gamma$ Boo) is one of ``the magnificent seven", the seven 
brightest $\delta$~Scuti stars which are on the target list of the 
BRITE microsatellite \citep{breger2008,kaiser2008}. The star is 
part of a multiple system of which the primary itself is found to be double 
by speckle interferometry with a separation of 0.069 arcsec 
\citep{hartkopf}.
The star was discovered as a short--period variable by 
\citet{guthnick1914}, with a pulsation amplitude of about 0.05~mag in $V$. 
The star pulsates with a dominant mode at $f$ = 21.28 c/d.

\citet{marilli1992} reported a possible detection of Ly--$\alpha$ emission in 
the star. From high--resolution spectra \citet{kennelly1992} were the first 
to detect a high--degree, non--radial pulsation in HD~127762. 
\citet{ventura2007} confirmed their result and found additional modes. 

Most elements appear to have solar abundance. The elements S, Y and Ba appear 
to be overabundant, but the analysis is based on just one or two lines.

\begin{figure}[ht]
\begin{center}
\resizebox{\hsize}{!}{\includegraphics{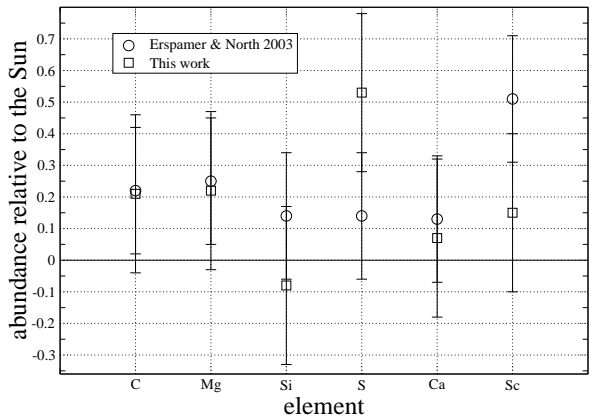}}
\caption{Comparison between the abundances in HD~127762 derived by
\citet{erspamer} and this work.}
\label{comp_hd127762}
\end{center}
\end{figure}

\citet{erspamer} derived the fundamental parameters and abundances for
HD~127762. They obtained \Teff\, = 7585, \logg\, = 3.74, \vmic\, = 3.1 and
\vsini\, = 123 and these parameters are in good agreement with our result. 
\citet{erspamer} measured abundances relative to the Sun using the solar
abundances by \citet{grevesse1998}. Figure~\ref{comp_hd127762} shows the
comparison between the abundances derived by \citet{erspamer}, after the
conversion to our adopted solar abundances, and this work. The abundances for
all the plotted elements are in agreement within the errors.
%
\section{Discussion}\label{discussion}
In the following Section we will compare the results of the abundance analysis 
of seven field $\delta$~Scuti stars with recent results for stars belonging to 
the Praesepe cluster \citep{fossati2007,fossati2008}. Furthermore, we will 
compare our results with results for four other field $\delta$~Scuti stars. 
Finally, we will discuss in detail the possible Am nature of the 
$\delta$~Scuti star HD 124953.

\subsection{Abundances in field and cluster $\delta$~Scuti stars}
\label{field vs. cluster}
\begin{figure}[ht]
\begin{center}
\resizebox{\hsize}{!}{\includegraphics{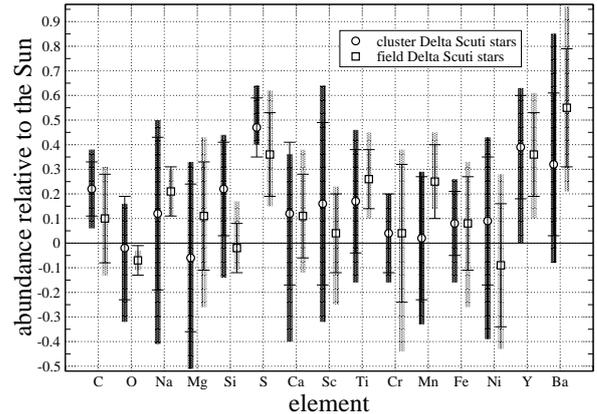}}
\caption{Mean abundance, relative to the Sun, and abundance range   
(shaded areas) of the sample of field $\delta$~Scuti stars (squares) and 
cluster $\delta$~Scuti stars (circles). The error bars 
correspond to the standard deviation of the mean abundance.}
\label{cluster.field}
\end{center}
\end{figure}

We compared two samples of $\delta$~Scuti stars (i) belonging to the 
Praesepe cluster, all with a common and well defined age 
\citep[$708 \pm 300$ Myr,][]{gonzalez}, 
and (ii) field $\delta$~Scuti stars. In Fig.~\ref{cluster.field} we show the 
mean abundance for each analysed element of the two samples 
and their abundance range (indicated by the dashed area). Note that the 
two samples of stars are
comparable since they are composed of almost the same number of stars (eight 
stars in the cluster sample and seven in the field sample) with similar
spectral type. 

The abundances of the two samples are close to solar with clear
overabundances of S, Y and Ba. 
In HD~124953 we find evidence for the star being an Am star, and we will 
discuss it separately in Sect.~\ref{disc-hd124953}. In the other stars we do 
not find abundance patterns typical for chemically peculiar stars.

The abundance patterns of the two samples appear to be quite similar.
To have a statistically more solid sample of stars we decided to 
merge the two samples. To test the validity of merging them we 
carried out a paired $t$--test. The test measures the probability that the 
samples are the same, and it was done separately for the 15 elements.
The $t$ values and corresponding probabilities that the two 
samples refer to the same statistical population are provided in 
Table~\ref{TAB t-test}. The bottom row of the table refers to the overall 
$t$--test merging all elements. The result, $t$ = 0.06, corresponds to a 
probability of 95\% that the two samples agree in their overall 
distribution of abundances. We consider this a sufficient argument for a common 
treatment of cluster and field $\delta$~Scuti stars.
\begin{table}
\caption{Statistical comparison of the abundances of 15 elements for cluster and field $\delta$~Scuti stars.}
\label{TAB t-test}
\scriptsize
\begin{center}
\begin{tabular}{rrrrrrrrrrr}
\hline
\hline
 &\multicolumn{3}{c}{cluster}& &\multicolumn{3}{c}{field}& & $t$ & Pr \\
El.&\# &mean   &rms   & &\# &mean   &rms   & &       &      \\
\hline
C  &$7$&$+0.22$&$0.28$& &$6$&$+0.10$&$0.30$& &$0.753$&$0.467$\\
O  &$4$&$-0.02$&$0.33$& &$2$&$-0.07$&$0.23$& &$0.196$&$0.854$\\
Na &$7$&$+0.12$&$0.41$& &$3$&$+0.21$&$0.25$& &$0.346$&$0.738$\\
Mg &$8$&$-0.06$&$0.40$& &$7$&$+0.11$&$0.33$& &$0.858$&$0.407$\\
Si &$8$&$+0.22$&$0.33$& &$7$&$-0.02$&$0.26$& &$1.511$&$0.155$\\
S  &$5$&$+0.47$&$0.29$& &$7$&$+0.36$&$0.30$& &$0.672$&$0.517$\\
Ca &$8$&$+0.12$&$0.40$& &$7$&$+0.11$&$0.29$& &$0.092$&$0.928$\\
Sc &$8$&$+0.16$&$0.43$& &$7$&$+0.04$&$0.29$& &$0.605$&$0.556$\\
Ti &$8$&$+0.17$&$0.34$& &$7$&$+0.26$&$0.27$& &$0.572$&$0.577$\\
Cr &$7$&$+0.04$&$0.31$& &$7$&$+0.04$&$0.37$& &$0.016$&$0.988$\\
Mn &$7$&$+0.02$&$0.37$& &$4$&$+0.25$&$0.28$& &$1.120$&$0.292$\\
Fe &$8$&$+0.08$&$0.30$& &$7$&$+0.08$&$0.31$& &$0.001$&$0.999$\\
Ni &$8$&$+0.09$&$0.37$& &$6$&$-0.09$&$0.34$& &$0.916$&$0.378$\\
Y  &$7$&$+0.39$&$0.34$& &$6$&$+0.36$&$0.30$& &$0.210$&$0.838$\\
Ba &$8$&$+0.32$&$0.39$& &$7$&$+0.55$&$0.34$& &$1.179$&$0.260$\\
\hline
   &$108$&     &      & &$90$&      &      & &$0.060$&$0.952$\\
\hline
\end{tabular}
\end{center}
\smallskip 

For each element, the number of stars, the mean abundance, and the standard 
error is given. The last two columns represent the $t$ values for a paired Student test and the 
resulting probabilities (Pr) that the two samples are taken from the same statistical population. The bottom row 
refers to the overall statistics for all elements.
\end{table}

\begin{figure}[ht]
\begin{center}
\resizebox{\hsize}{!}{\includegraphics{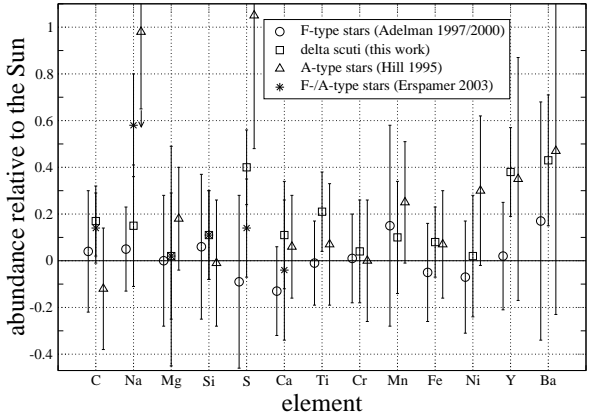}}
\caption{Comparison of the mean abundances and standard deviations 
for four samples of stars. The Na abundance of the A--type stars
\citep{hill1995} is an upper limit.}
\label{comp.normal}
\end{center}
\end{figure}

In order to be able to confirm that $\delta$~Scuti stars in general have
abundances that are typical of non--variable early F-- and late A--type 
stars we have used three comparison samples. We took a sample 
of non--variable late F--type stars (5 stars, \Teff\, $\sim$ 6800~K), 
analysed by \citet{adelman1997,adelman2000}, non--variable early F-- 
and late A--type stars (with a mean \Teff\, corresponding to our sample; 11 
stars, \Teff\, $\sim$ 7600~K) analysed by \citet{erspamer} and 
of non--variable early A--type stars (15 stars, \Teff\, $\sim$ 9300~K), 
analysed by \citet{hill1995}. In Fig.~\ref{comp.normal} we compare the mean 
abundances and uncertainties for the four samples. For most elements the 
abundances are close to solar, although we find slightly higher abundances 
of S, Y and Ba. In our sample we find a much lower abundance of Na compared 
to the sample from \citet{erspamer}. This is probably due to the fact that 
the Na lines used by them (\ion{Na}{i} doublet at $\sim$5889 
and $\sim$5895 \AA) to derive the Na abundance are known to be affected by 
strong \nlte\, effects that would lead to a higher abundance. We avoided 
the use of these two lines.
%
\subsection{Comparison with single $\delta$~Scuti stars}\label{comparison}
\begin{figure}[ht]
\begin{center}
\resizebox{\hsize}{!}{\includegraphics{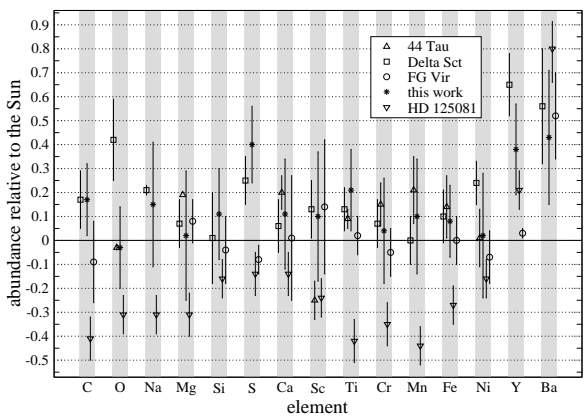}}
\caption{Comparison of the mean abundance pattern of our sample with four 
other $\delta$~Scuti stars: FG~Vir, $\delta$~Sct, 44~Tau and HD~125081. The 
adopted error bars for the single stars are taken from the original papers.}
\label{comp_other}
\end{center}
\end{figure}

We compared the abundance pattern in our sample with results for four 
single field $\delta$~Scuti stars: FG~Vir \citep{mittermayer2003}, 
$\delta$~Sct \citep{yushenko2005}, 44~Tau \citep{zima2007} and HD~125081 
\citep{bruntt}. In Fig.~\ref{comp_other} we compare the abundances of these 
four objects.

The mean effective temperature of our sample is 7600~K. The comparison stars 
are slightly cooler: HD~125081 has \Teff\, = 6400~K, $\delta$~Sct and 44~Tau 
have \Teff\, = 7000~K, and FG~Vir has \Teff = 7425~K. The results presented in 
Fig.~\ref{comp.normal} and \ref{comp_other} show that there is no obvious 
temperature effect, although the coolest star, HD~125081, has low abundance
of both light and Fe--peak elements. To be able to confirm that this is a 
temperature effect, more comparison stars would be needed.

\subsection{HD~124953: an Am--type $\delta$~Scuti star?}\label{disc-hd124953}
HD~124953 was first classified as a metallic--line star by \citet{walker1966},
based on photometric indices. This has been confirmed by several other studies
\citep[e.g. ][]{hauck}.
An abundance analysis to give a definite classification as an Am star was 
never performed. 
\citet{bertaud1967}, however, concluded it was a normal F0IV star 
in which the \ion{Ca}{i} lines are weaker than usual.

It is well--established that Am stars show underabundances of C, N, O, Ca and 
Sc and overabundances of the Fe--peak elements, Y, Ba and of the rare earths 
elements \citep{adelman1997, fossati2007}.

\begin{figure}[ht]
\begin{center}
\resizebox{\hsize}{!}{\includegraphics{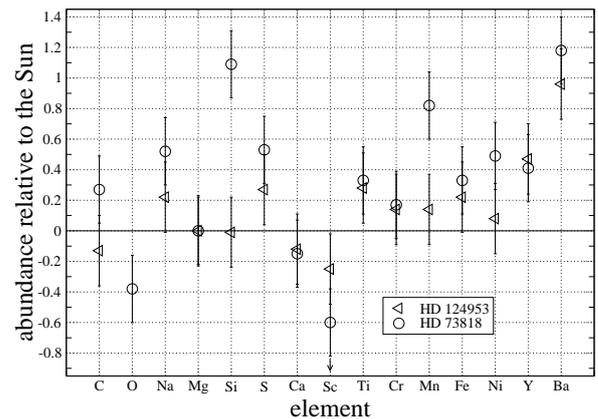}}
\caption{Abundances of HD~124953 are compared with HD~73818. The latter is a 
fast--rotating Am star belonging to the Praesepe cluster.}
\label{hd124953_abn}
\end{center}
\end{figure}

\citet{charbonneau} gave a rotational velocity limit of 90~\kms\, above which 
diffusion processes cannot cause Am peculiarities. Within an error of 5\% the 
\vsini\, of HD~124953 is close to this limit, but certainly below 90 \kms. 
If the star is viewed equator on, the rotational velocity 
would be low enough to allow the star to be a mild Am. \citet{fossati2008} have
confirmed observationally that the abundances of the elements characteristic 
for Am stars are strongly related to the rotational velocity. In 
Fig.~\ref{hd124953_abn} we compare the abundances of HD~124953 with those 
obtained for the fastest rotator analysed by \citet{fossati2007}: HD~73818. 
This Am star has a \vsini\, of 66 \kms\, and an effective temperature 
(\Teff\, = 7230~K) comparable to HD~124953. The Am star HD~73818 is a 
member of the Praesepe cluster.

\begin{figure}[ht]
\begin{center}
\resizebox{\hsize}{!}{\includegraphics{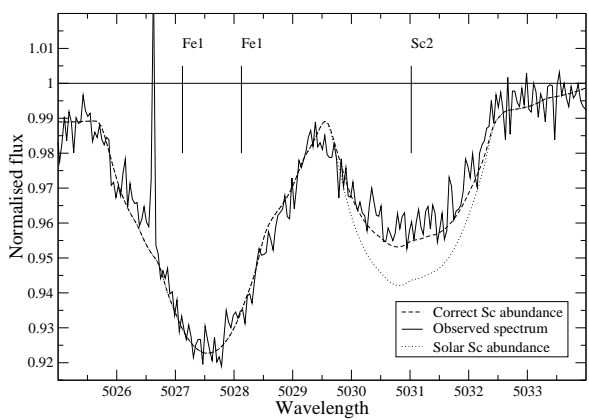}}
\caption{Observed spectrum of HD~124953 and synthetic spectra around 
the \ion{Sc}{ii} line at 5031.2~\AA. The dotted line is for solar abundance 
of Sc and the dashed line is for the derived abundance.}
\label{am}
\end{center}
\end{figure}

HD~124953 shows underabundances of C, Ca, Sc and overabundances of the Fe--peak
elements, Y and Ba, which is typical for Am--type stars. Note that the 
high abundance of Si in HD~73818 was not seen in the other Am stars in the 
Praesepe sample \citep[see][]{fossati2007}. The only relevant difference 
between the two objects is the Sc abundance. According to \citet{fossati2008} 
the Sc underabundance increases with the rotational velocity in contrast with 
predictions from diffusion models \citep[see ][]{talon2006,alecian2008}. 
If Sc behaved like the other underabundant elements, as predicted by 
diffusion models, its abundance should be close to the relative abundance 
of Ca. This is the case for HD~124953, but not for HD~73818. Figure~\ref{am}
shows the observed and synthesised \ion{Sc}{ii} line at 5031.021~\AA\, with
solar abundance (dotted line) and the derived Sc abundance (dashed line). It is
clear that the solar Sc abundance is too high. The synthetic spectrum for 
the fitted abundance is shown in Fig.~\ref{am} and it is 0.43~dex less than 
the solar value.

The diffusion processes that lead to chemical peculiarities inhibit 
$\delta$~Scuti type pulsation \citep[see ][ for Ap and Am stars]{kurtz2000}. 
However, there are some important exceptions. Classical Am stars that have 
been proven to be low--amplitude $\delta$~Scuti stars do exist, such as 
HD~1097 \citep{kurtz1989}, HD~13038 \citep{martinez1999a}, and HD~13079 
\citep{martinez1999b}. Recently, more Ap and Am stars showing pulsation 
were reported by \citet{joshi2006}. \citet{henry2005} and \citet{king2008} 
reported on the discoveries of Am stars showing hybrid pulsations, both in 
the $\gamma$~Doradus and the $\delta$~Scuti pulsation period ranges. 

The Am character of a star can be derived from the metallicity index 
($\delta$$m_1$) and luminosity index ($\delta$$c_1$), as was done for HD~124953. 
Both indices are sensitive to the strong line--blanketing found in the Am and 
Ap stars. The analysis here presented for HD~124953 has confirmed its 
photometric classification as Am star. This is probably the 
first abundance analysis of a pulsating Am star. 

Nearly all Am stars are found to occur in binary systems with orbital periods 
between 1 and 10 days in which the rotation and orbital periods are locked.
As a result of the relatively slow rotation, chemical peculiarities can be 
found in these stars \citep{kurtz2000}. For a supposedly pulsating Am star in 
a binary, it is therefore important to prove that the pulsating star is 
really the Am 
star and not the other component in the system. A historic example is 32~Vir, 
for which claims of $\delta$~Scuti pulsation in Am stars were dismissed by 
\citet{kurtz1976}. Given the binary nature of the star, they found no 
convincing evidence for the pulsation of a single classical Am star. Later it 
was found that the primary is in fact a pulsating $\rho$~Puppis star 
\citep{mitton1997}.

HD~124953 is marked as a suspected spectroscopic binary (``SB?") in the Bright 
Star Catalogue \citep{hoffleit1964}. In our spectrum of the star, however, we 
see no evidence of a possible companion star, as already mentioned in
Sect.~\ref{observations}. The radial velocity we find for the star 
(5.3$\pm$4.0~\kms) is in agreement with those found in the literature (3 
references in SIMBAD), and shows no evidence of any significant radial 
velocity variation. Hence, at this point we have no evidence for the binary 
nature of the star, and it remains a candidate as an Am star showing 
$\delta$~Scuti pulsations.
\section{Conclusions}\label{conclusions}
We obtained high--resolution, high signal--to--noise ratio spectra of 
seven bright field $\delta$~Scuti stars. Due to their brightness, these stars 
have in general not been studied in great detail photometrically. Hence, 
time--series photometry to study the pulsation frequencies and/or measure the 
phase lag in different colors for mode identification has not been carried out. 
The analysed stars are potential future targets for microsatellite projects 
such as BRITE \citep{kaiser2008}, and we discussed briefly their 
pulsation properties.

We compared the abundance pattern of the seven field $\delta$~Scuti stars 
with a sample of $\delta$~Scuti stars belonging to the Praesepe cluster 
\citep{fossati2007,fossati2008}. A $t$--test confirms that the 
abundance patterns of the two samples (field and cluster 
$\delta$~Scuti stars) are comparable, allowing us to build one homogeneous 
sample of fifteen early F-- and late A--type $\delta$~Scuti stars.

We used the abundances of non--variable F-- and A--type stars 
published by \citet{adelman1997,adelman2000}, \citet{erspamer} and 
\citet{hill1995}, to generate a typical abundance pattern for stars of similar 
spectral types. From a direct comparison, we showed that the mean 
abundance pattern of our sample of $\delta$~Scuti stars and of the 
non--variable F-- and A--type stars are comparable. The sample that we 
built allows us to conclude that generally the abundances of $\delta$~Scuti 
stars are comparable to those of normal stars of similar spectral types. We 
also compared the abundance pattern in our sample with results for four 
single field $\delta$~Scuti stars from the literature, and found good general 
agreement.

HD~124953 has previously been classified as a metallic--line star based on 
photometric indices. Our abundance analysis confirms this classification. It 
would be worthwhile to follow up on this star with photometric and 
spectroscopic data to better constrain the pulsation characteristics and the 
possible binarity of which there is no evidence from our spectrum. In case of 
confirmed binarity it would be necessary to prove that the Am star pulsates 
and not the secondary component.
\begin{acknowledgements}
We thank Michel Breger for pleasant and fruitful discussions and James Silvester
for the english revision of the manuscript. We gratefully acknowledge the 
referee Hans Bruntt for useful suggestions. LF has received support from the 
Austrian Science Foundation (FWF project P17890--N2). KK acknowledges funding 
through the Austrian Science Foundation (FWF project T395). This paper is 
based on observations obtained using the SOPHIE spectrograph at the 
Observatoire de Haute--Provence (France). We acknowledge also the OPTICON 
program (Ref number: 2007/011) for financial support given to the 
observing run.
\end{acknowledgements}
\Online

\end{document}